\documentclass[preprint,12pt]{elsarticle}

\usepackage{amssymb}
\usepackage{amsmath}
\usepackage{ragged2e}
\usepackage{subcaption}
\usepackage{array} 

\usepackage{float}

\begin{document}

\begin{frontmatter}

\vspace*{0.2in}

\begin{flushleft}
{\Large
\textbf\newline{Exploring the core-periphery and community structure in the financial networks through random matrix theory} 
}
\newline
\\
Pawanesh \textsuperscript{1,\dag,*},
Imran Ansari\textsuperscript{1,\dag,*}
Niteesh Sahni\textsuperscript{1}
\\
\bigskip
\textsuperscript{\textbf{1}} Department of Mathematics, Shiv Nadar Institution of Eminence Deemed to be University, Delhi-NCR-201314, India
\\
\bigskip

%
%
\dag\ Corresponding authors: py506@snu.edu.in, ia717@snu.edu.in\\
* These authors contributed equally to this work.







\end{flushleft}

\makeatletter
\renewcommand\@seccntformat[1]{\csname the#1\endcsname.\quad}
\makeatother
\begin{abstract}

In finance, Random Matrix Theory (RMT) is an important tool for filtering out noise from large datasets, revealing true correlations among stocks, enhancing risk management and portfolio optimization. In this study, we use RMT to filter out noise from the full cross-correlation matrix of stock price returns for the NIFTY 200 and NIFTY 500 indices on the National Stock Exchange of India. In addition, we applied network theory tools to analyze market and sector modes as filtered correlation structures to study local interactions within financial networks. This allows us to study the very fundamental properties of networks, such as the core-periphery and the community structure of constructed networks over these filtered modes, and compare the results with the network constructed over the full cross-correlation matrix. The results suggest that the core-periphery structure is contained in the market mode, while the community structure is in the sector mode. Thus, both modes outperform the full cross-correlation in terms of capturing the essential respective structure of the network. Furthermore, we used these insights to build portfolios based on communities of the networks corresponding to the sector mode and the network corresponding to the full cross-correlation matrix. The results suggest that the portfolio constructed on the complete cross-correlation-based matrix performs better than the sector mode. These insights provide a greater understanding of RMT application in the financial market.
\end{abstract}




\end{frontmatter}



\section{Introduction}

\begin{justify}

There are a huge number of time series datasets available in the financial market, such as price and trading volume series, which has led to a lot of research on financial correlation-based networks. A common practice is the construction of networks based on the correlation between these time series datasets. Over the past few years, researchers have frequently employed some popular information filtering approaches in network development such as MST, PMFG \cite{mantegna1999hierarchical,tumminello2005tool,tumminello2007correlation,pozzi2007dynamical,song2011evolution,eryigit2009network,matteo2010use,aste2010correlation}, and threshold algorithms \cite{Yang2008, Tse2010, Huang2009, Onnela2003, Kim2007, Junior2013, Nobi2014, Nobi2014a, Boginski2006, Namaki2011, Boginski2005, Heiberger2014}.  On the other hand, many researchers have used Random Matrix Theory for filtering noise from the empirical correlation matrix to understand the co-movement of the stocks based on daily logarithmic return times series data and also proposed that the correlation matrix could be decomposed into three modes market mode, sector mode, and Noise mode such that market mode captures the global structure and sector mode contains the community structure of the financial market \cite{gopikrishnan2001quantifying,pan2007collective,macmahon2015community,jiang2014structure}.  In 2001, a comprehensive analysis was performed using Random Matrix Theory (RMT) on the cross-correlation matrix of daily stock returns in relation to an established market such as the United States \cite{gopikrishnan2001quantifying}. Later, in 2007, Pan and Sinha \cite{pan2007collective} analyzed the cross-correlation matrix of daily stock returns in the US and India, two developed and emerging nations. The spectrum properties of the empirical correlation matrix of the rate of return were compared with the random matrix that would have emerged in the event that the returns of the shares were not correlated \cite{gopikrishnan2001quantifying,pan2007collective}.  Pan and Sinha also proposed a method for selecting the optimal threshold of eigenvalues that falls into the sector mode. A sufficient amount of work has been done by combining the RMT and network filtration approaches simultaneously to understand the correlation-based networks of the financial systems. In 2014, Jiang et al. \cite{jiang2014structure} comprehensively studied the SSE and NYSE markets by building the filtered network over the RMT sector mode and investigated the interaction structure of communities in financial markets and revealed that local interactions between the business sectors (sub-sectors) are mainly contained in the sector mode.

In 2015, MacMahon and Garlaschelli \cite{macmahon2015community} studied the market mode, sector mode, and both modes altogether and showed the non-obvious correlations between stocks from different industry sectors. In particular, the identification of internally correlated and mutually anti-correlated communities is crucial for portfolio optimization and risk management. They also assess the stability of these communities over varying time scales, identifying "hard stocks" that consistently remain within the same community and "soft stocks" that shift between communities. In 2021, Purqon and Jamaludin \cite{purqon2021community} introduced two hybrid methods, which are based on RMT, to detect and analyze dynamic complex networks and showed the effectiveness of these methods in identifying detailed community structures and uncovering hidden interactions within the market, highlighting their unique strengths in modularity and network decomposition.

According to the literature discussed above, random matrix theory (RMT) offers a powerful framework for distinguishing meaningful correlations from random noise, thus providing a clearer view of the system's structure. On the other hand, in network science, just like community structure, another very important feature of the network is core-periphery structure, which has been very hot research in the current decade to study real-world systems like transportation, finance, communication networks, etc.

Although the previous literature has done little work on community detection using RMT decomposition, here we did a comprehensive dynamic analysis of community structures over time and further used the insights to propose a portfolio selection strategy based on community information as mentioned in \cite{zhao2021community}. Similarly, the core-periphery structure, another important mesoscale structure, remains relatively unexplored using RMT decomposition. Therefore, we also conducted an in-depth analysis of the core-periphery structure using the RMT approach. The detection of core-periphery structures by RMT is a novel contribution to network science and deserves recognition in the field.


%


In network science, the core-periphery and the community both the structures have been fundamental organizational patterns. In the core-periphery structure, the core nodes are densely connected with each other and some periphery nodes, whereas periphery nodes have sparse connections with one another, while the communities are common structural entities in the networks, usually with a higher internal and a lower external link density. During the past few decades, these structures have been applied to various types of networks, including social networks \cite{Borgatti1999,Rossa2013,holme2005core,Zhang2015,kojaku2017finding}, protein-protein interaction (PPI) networks \cite{Rossa2013,yang2014overlapping}, financial networks \cite{in2020formation,craig2014interbank}, transportation networks \cite{Rombach2017,Rossa2013,kojaku2017finding}, and neural networks \cite{Rossa2013,tunc2015}. A range of methods has been developed to detect and analyse for core-periphery  \cite{Rossa2013,holme2005core,surprise2018,Rombach2017,in2020formation,Borgatti1999} and the community structure \cite{blondel2008fast,newman2004finding,raghavan2007near} in different domains.

In this paper, we focus on two important aspects of networks: the core-periphery structure and the community structure. We study these structures using the random matrix theory (RMT) approach in two financial data sets, the NIFTY 200 and NIFTY 500 indices, collected from the Indian National Stock Exchange during the period from January 2010 to December 2022. By applying RMT, we filter out noise and identify significant eigenvalues and eigenvectors, which help isolate the true correlation structure. This refined correlation matrix allows us to analyze the hierarchical organization of the network, including the identification of core-periphery and community structures. Thus, the advanced core-periphery and community detection algorithms uncover these structures, providing deeper insights into connectivity patterns and relationships of the financial system. Our analysis shows that the correlation matrix corresponding to the largest eigenvalue, known as the market mode, captures a more significant core-periphery structure compared to the full and sector mode correlation matrix. Moreover, the correlation matrices corresponding to a few of the largest eigenvalues (three in the case of NIFTY 200 and five in the case of NIFTY 500), referred to as the Sector mode, capture a more significant community structure than the market and the full cross-correlation matrix. As an application, we also demonstrate that communities can serve as effective indicators for risk mitigation and return optimization. Following this reference \cite{zhao2021community}, we show that investing in inter-community assets not only reduces risk but also yields better returns compared to the market portfolio. These results are consistent with both the uniform and Markowitz methods.

The remaining part of the paper is structured into 5 sections. In Section 2, we describe the data used in our analysis. In Section 3, we provide a brief discussion of methods utilized in our study. In Section 4, we give a detailed discussion of the findings and results of this study. Lastly, in the final section, we summarize our key observations and future implications of the work.

\end{justify}

\section{Data description}
\begin{justify}
We utilized two datasets from the National Stock Exchange (NSE) of India, namely NIFTY 500 and NIFTY 200. The NIFTY 500 dataset consists of daily closing prices for all 500 stocks listed on the NIFTY 500 index spanning from January 1, 2010, to December 31, 2022. 

Due to incomplete price data, we excluded some stocks from our analysis, resulting in a refined data set comprising 312 stocks and 3204 trading days with complete information. To analyze the returns, we computed the daily logarithmic returns defined as
\[
r_{i}(t) = \ln p_{i}(t+1) - \ln p_{i}(t)
\]
where \( p_{i,t} \) represents the closing price of the stock \( i^{th} \) on day \( t \).

In calculating these returns, we omitted data corresponding to non-successive trading days. For example, if a holiday occurred between Friday and Monday, making the days non-consecutive, the return for Monday was excluded from the analysis. Taking into account such adjustments, our final data set consists of 3204 daily logarithmic return values for each of the 312 stocks.

Similarly, on the other hand, for NIFTY 200, we get 140 stocks after filtration with 3208 trading days.
\end{justify}
\section{Methods}
\begin{justify}
In this section, we first introduce Random Matrix Theory (RMT), followed by an overview of several well-known core-periphery algorithms, including Rossa \cite{Rossa2013}, Rombach \cite{Rombach2017}, and MINRES \cite{boyd2010computing}, used to identify and analyze core-periphery structures within networks. In addition, we will also employ several well-established community detection algorithms, such as the Louvain Algorithm \cite{blondel2008fast}, label propagation \cite{raghavan2007near}, and Newman Girvan \cite{newman2004finding}. Finally, we will compare network partitions using some well-known measures such as Normalized Mutual Information (NMI) \cite{danon2005comparing,danon2006effect}.

\subsection{Random matrix theory}

Random matrix theory is a statistical framework that is widely used in diverse domains such as finance, wireless communications, and complex systems. To understand the behavior and distribution of large matrices, it concentrates on their eigenvectors and eigenvalues. 

Let $r_{i}(t)$ be the daily logarithmic price return at time $t$ of the indices $i = 1,2,3,...,N$ where $N$ is the total number of stocks. The time spans $t = 1,2,3,...T$ where $T$ is the maximum time of the data set.

\

The complete cross-correlation matrix $C=[C_{ij}]$ is expressed in terms of $r_{i}(t)$ as:
$$C_{ij} =  \frac{E(r_{i}r_{j})-E(r_{i})E(r_{j})}{\sqrt{E(r_{i}^2)-E(r_{i})^2}\sqrt{E(r_{j}^2)-E(r_{j})^2}} $$

Where $E(r_{i})$ represents the expected value of returns corresponding to the $i$th stock. Here $C^{f}$ is a real symmetric matrix with $C_{ij}=1$, and all the values of $C_{ij}$ are in the interval $-1 \leq C_{ij}\leq 1$. The statistical properties of the correlation matrix are well established in the literature \cite{mantegna1999hierarchical, tumminello2007correlation, tumminello2005tool, pozzi2007dynamical, song2011evolution, eryigit2009network, matteo2010use, aste2010correlation, Yang2008, Tse2010, Huang2009, Onnela2003, Kim2007, Junior2013, Nobi2014, Nobi2014a, Boginski2006, Namaki2011, Boginski2005, Heiberger2014, gopikrishnan2001quantifying, jiang2014structure, macmahon2015community, purqon2021community,pan2007collective}. Especially, when there are a large number of variables $N$ and large sample points $T$ and under the hypothesis that the correlation matrix is random, then the distribution of eigenvalues of $C$ is approximated by the Marchenko-Pastur distribution. Here, we compare the properties of the correlation matrix $C$ with those of the random correlation matrix (Wishart matrix) \cite{baker1998random}.  \\

For large $N \to \infty$ and $T \to \infty$ such that $Q = \frac{T}{N} \geq 1$, the probability density function of the eigenvalues $\lambda$ of the random correlation matrix are given as follows;

\[
P_{rm}(\lambda) =
\begin{cases}
\frac{Q\sqrt{(\lambda_{max}-\lambda)(\lambda_{min}-\lambda)}}{2\pi \lambda} & \text{if } \lambda_{min} \leq \lambda \leq \lambda_{max}, \\
0 & \text{otherwise}.
\end{cases}
\]

where $\lambda_{min}$ and $\lambda_{min}$ are the maximum and minimum eigenvalues of $C$ given by
\[
\lambda_{\text{min}} = \left(1 - \sqrt{\frac{Q}{2}}\right)^2 \quad \text{and} \quad \lambda_{\text{max}} = \left(1 + \sqrt{\frac{Q}{2}}\right)^2
\]

The complete cross-correlations \(C\) can be decomposed into different eigenmodes: \(C_{ij} = \sum_{\alpha=0}^{N-1} C^{\alpha}_{ij}, \quad C^{\alpha}_{ij} = \lambda_{\alpha} u^{\alpha}_{i} u^{\alpha}_{j}\), where \(\lambda_{\alpha}\) is the \(\alpha\)-th eigenvalue of \(C_{ij}\), and \(u^{\alpha}_{i}\) is the \(i\) -th component of the \(\alpha\) -th eigenvector. We define the full cross-correlation matrix into three modes, namely market mode, sector mode, and random mode, as follows. The market mode \(C^{\text{M}}\): \(C^{\text{M}}_{ij} = C^{0}_{ij}\), corresponding to the largest eigenvalue of \(C_{ij}\); the sector mode \(C^{\text{S}}\): \(C^{\text{S}}_{ij} = \sum_{\alpha=1}^{k} C^{\alpha}_{ij}\), where \(k\) is the number of eigenvalues, other than the largest eigenvalue, that deviate from the \([\lambda_{\min}, \lambda_{\max}]\) interval; the random mode \(C^{\text{R}}\): \(C^{\text{R}}_{ij} = \sum_{\alpha=k+1}^{N-1} C^{\alpha}_{ij}\). 
Here, we choose the method described in reference \cite{pan2007collective} to decide the threshold of eigenvalues that falls in sector mode over the entire 13 years of data sets. We find that in the sector mode five largest eigenvalues (excluding the largest one) fall for the NIFTY 500 dataset, and similarly three eigenvalues (excluding the largest one) fall for the NIFTY 200 dataset.

Further, we construct the Planar maximally filtered network over the complete cross-correlation matrix $C$, market mode $C^{M}$ and sector mode $C^{S}$. Then the adjacency matrix of these PMFG networks is denoted as \(C^{f}\), \(C^{\text{mar}}\), and \(C^{\text{sec}}\), respectively.


\subsection{Core-periphery detection in networks}

In this section, we provide a brief overview of well-established methods used to detect the core-periphery structure in networks.

\subsubsection{Rossa method}
Rossa et al.~\cite{rossa2013profiling}  approach core-periphery detection from a random walk Markov chain perspective. In this model, the vertices of the network are states of the Markov chain, and the transition probability matrix \((p_{ij})\) is defined as \(p_{ij} = \frac{a_{ij}}{\sum_h a_{ih}}\).

Let \( S \) be a subgraph of a graph \( G \). The persistent probability \( \alpha_S \) measures the cohesiveness of \( S \) and is defined as

\[
\alpha_S = \frac{\sum_{i,j \in S} \pi_i p_{ij}}{\sum_{i \in S} \pi_i} = \frac{\sum_{i,j \in S} a_{ij}}{\sum_{i \in S, j \in V(G)} a_{ij}},
\]

where \( \pi_i > 0 \) is the stationary probability of being at node \( i \), given by \( \pi_i = \frac{C_D(i)}{\sum_{j \in V(G)} C_D(j)} \), and \( C_D(i) = \sum_h a_{ih} \) represents the weighted degree of the \(i\)-th vertex.

To determine the coreness value for each vertex, start with the vertex of the least weighted degree, denoted \( S_1 \). Assume \( S_1 = \{1\} \) and set \( \alpha_1 \equiv \alpha_{S_1} = 0 \). For the next step, consider subsets \( S_2^{(j)} \equiv S_1 \cup \{j\} \) for \( 2 \leq j \leq N \) and compute \( \alpha_{S_2^{(j)}} \) for each. Select \( S_2 \) as the subset \( S_2^{(k)} \) with the minimum \( \alpha_{S_2^{(k)}} \) and set \( \alpha_2 \equiv \alpha_{S_2} \), which is the coreness value for vertex \( k \). Note that \( S_2 \) has the smallest persistence probability and \( \alpha_1 \leq \alpha_2 \). Continue this process to construct \( S_3 \) from \( S_2 \) and compute \( \alpha_3 \), ensuring \( \alpha_1 \leq \alpha_2 \leq \alpha_3 \). Repeat until all vertices are assigned a coreness value.\\
\subsubsection{MINRES method}
Boyd et al. \cite{boyd2010computing} aims to find a MINRES vector \( \mathbf{c} \) such that the adjacency matrix \( \mathbf{A} \) is approximated by \( \mathbf{c}\mathbf{c}^T \). This approximation minimizes the sum of squared differences off the diagonal. Specifically, it seeks to find a vector \( \mathbf{c} \) that minimizes

\[
\sum_{i} \sum_{j \neq i} [a_{ij} - c_i c_j]^2.
\]

By taking the partial derivative with respect to each element of \( \mathbf{c} \), we get

\[
c_i = \frac{\sum_{j \neq i} a_{ij} c_j}{\sum_{j \neq i} c_j^2},
\]

which leads to an iterative process for computing the MINRES vector.\\
\subsubsection{Rombach method}
Rombach et al.~\cite{rombach2017core} assign core scores to vertices in weighted, undirected networks through an optimization process. The goal is to maximize the quality function.

\[
Q(\alpha, \beta) = \sum_{i=1}^N \sum_{j=1}^N a_{ij} c_i(\alpha, \beta) c_j(\alpha, \beta),
\]

where \(c_i^*(\alpha, \beta)\) is initialized as:

\[
c_i^*(\alpha, \beta) = \begin{cases} 
\frac{i(1-\alpha)}{2 \lfloor \beta N \rfloor}, & \text{if } i \in \{1, \ldots, \lfloor \beta N \rfloor\} \\
\frac{(i - \lfloor \beta N \rfloor)(1-\alpha)}{2(N - \lfloor \beta N \rfloor)} + \frac{1+\alpha}{2}, & \text{if } i \in \{\lfloor \beta N \rfloor + 1, \ldots, N\}.
\end{cases}
\]

The parameters \(\alpha\) and \(\beta\) range between 0 and 1. The optimal permutation of \((c_i^*)\) that maximizes \(Q\) is denoted \((c_i)\). The core score (CS) for each vertex \(i\) is:

\[
CS(i) = Z \sum_{(\alpha, \beta)} c_i(\alpha, \beta) Q(\alpha, \beta),
\]

where \(Z\) normalizes \(\max_{1 \leq j \leq N} CS(j) = 1\). This study uses 10,000 pairs \((\alpha, \beta)\) uniformly sampled from \([0, 1] \times [0, 1]\).

\subsection{Finding and evaluating communities}

Communities, which are known as groups, modules, or clusters, are structural components that are often found in complex networks. They have a higher internal link density than an exterior one, and there is not yet a universally recognized definition. In the literature, many well-established methods are proposed to find the communities in the network \cite{fortunato2010community,fortunato2016community,cherifi2019community}. First, we will introduce modularity \cite{newman2004finding}, a measure of the strength of the division of a network into communities. Then we will provide a brief description of each of the community detection methods used in this paper.
In mathematical terms, let us denote by $c_i$ the class or type of vertex $i$, which is an integer $1 \ldots n_c$, with $n_c$ being the total number of classes. Then the difference between the total number of edges in a community and the total expected number of edges between the vertices $i$ and the vertex $j$ of the same community can be expressed as follows;

\[
\frac{1}{2} \sum_{i,j} ^{N}\left( A_{ij} - \frac{k_i k_j}{2m} \right) \delta(c_i, c_j)
\]

Where $N$ is the number of nodes and $A_{ij}$ is the element of the adjacency matrix ($A_{ij}=A_{ij}=1)$. $\delta(c_{i}, c_{j})$ is the Kronecker delta, which is equal to 1 when $i$ and $j$ are in the same community; else 0, conventionally one calculates not the number of such edges, but the fraction, which is given by the same expression divided by the number $m$ of edges:

\[
Q = \frac{1}{2m} \sum_{i,j} \left( A_{ij} - \frac{k_i k_j}{2m} \right) \delta(c_i, c_j)
\]

This quantity $Q$ is called modularity \cite{newman2003mixing,newman2004finding} and is a measure of the extent to which the vertices are connected in the network. It is always $Q < 1$. It is $Q > 0$ if there are more edges between vertices of the same type than we would expect by chance, and $Q < 0$ if there are fewer. The Modularity for a weighted network \cite{newman2004analysis} can be given as follows;

\[
Q = \frac{1}{2M} \sum_{i,j} \left( w_{ij} - \frac{s_i s_j}{2m} \right) \delta(c_i, c_j)
\]

where the number of edges $m$ is replaced by $M = \frac{1}{2} \sum_{i=1}^N \sum_{j=1}^N w_{ij}$ (with $w_{ij}$ denoting the weight of the edge between the nodes $i$ and $j$), and the degree of the node is replaced by the node strength defined, for example, for node $i$ as $s_i = \sum_{l=1}^N w_{li}$.
\subsubsection{Louvain algorithm}

The Louvain \cite{blondel2008fast} is a Greedy algorithm for community detection with $o(nlogn)$ run-time. This algorithm is divided into two phases that are repeated iteratively. It greedily maximizes modularity $Q$. 
In phase one, each node is assigned to a unique community. Then choose a random node \(i\) and merge it with a community of a neighbouring node \(j\). Next, compute the modularity gain \(\Delta Q\) and allow this merging if \(\Delta Q\) is positive. This process repeats until no further improvement is possible, reaching a local maxima of modularity. The modularity gain \(\Delta Q\) by moving an isolated node \(i\) to a community, let say \(U\) is:
\[
\Delta Q = \left[ \frac{{\sum_{\text{in}} + k_{i,\text{in}}}}{{2m}} - \left( \frac{{\sum_{\text{tot}} + k_i}}{{2m}} \right)^2 \right] - \left[ \frac{{\sum_{\text{in}}}}{{2m}} - \left( \frac{{\sum_{\text{tot}}}}{{2m}} \right)^2 - \left( \frac{{k_i}}{{2m}} \right)^2 \right]
\]
where \(\sum_{\text{in}}\) is the sum of weights of links inside \(U\), \(\sum_{\text{tot}}\) is the sum of weights of links incident to nodes in \(U\), \(k_i\) is the sum of weights of links incident to \(i\), \(k_{i,\text{in}}\) is the sum of weights of links from \(i\) to nodes in \(U\), and \(m\) is the sum of weights of all links in the network. The second phase constructs a new weighted network with self-loops. The communities obtained in the first phase are contracted into aggregated communities, and the network is created accordingly. Two aggregated communities are connected if there is at least one edge between the nodes of the corresponding communities. The weight of the edge between the two communities, for example $U$ and $V$ is $W_{UV}= \sum_{i\in U} \sum_{j\in V} A_{ij}$ which is the sum of the weights of all edges between their corresponding communities, and the weights of the self-loops for an aggregated community $U$ is given as $W_{UU}=\sum_{i\in U} \sum_{j\in U} A_{ij}$. For the precise details, the reader may refer to the original paper of the Louvain algorithm \cite{blondel2008fast}.

\subsubsection{Newman girvan algorithm}

The Girvan-Newman algorithm \cite{newman2004finding} is a widely recognized technique for community detection in networks developed in 2002. This algorithm is a divisive rather than an agglomerate. The detected communities involve progressively removing edges with the highest centrality of betweenness. Betweenness centrality for an edge is the number of shortest paths that pass through it, making it a critical measure for identifying edges that serve as bridges between communities.
The algorithm's steps for community detection are summarized below; 

\begin{itemize}
    \item  Calculate the betweenness for all edges in the network.
    \item  Remove the edge with the highest betweenness centrality.
    \item  Recalculate edge betweenness centrality for the remaining edges in the network.
    \item  Steps 2 and 3 are repeated until no edges remain.
\end{itemize}

This approach effectively detects communities by successively peeling away the weakest edges that connect different groups. The quality of the detected communities is evaluated using a measure called modularity (Q). The Girvan-Newman algorithm is particularly useful for networks where community structures are not immediately obvious and where traditional methods might fail to detect overlapping or hierarchical communities. However, it can be computationally intensive, especially for large networks, due to the repeated calculation of edge betweenness. Despite this, the algorithm remains a foundational technique in the study of network community detection.

\subsubsection{Asynchronous label propagation algorithm}

The asynchronous label propagation algorithm (ALPA) \cite{raghavan2007near} models the diffusion of labels across the network's connections, with each node assigned a community label. Applying a majority rule, these labels are updated continuously based on the nearby nodes. As the labels propagate, densely connected nodes tend to converge on a common label, creating distinct communities. Importantly, this algorithm does not have any predefined metric or objective to optimize. first, each node is assigned a unique community label. The algorithm then iteratively performs the following asynchronous update process until each node has more neighbors within its community than in any other.
\begin{itemize}
    \item  The nodes are randomly ordered.
    \item  Each node's label is updated one by one according to the labels of its neighbors, in the prescribed order. Importantly, a node takes the label of the community to which most of its neighbors currently belong. The neighbors of a node might have already modified their labels during an iteration. The influence of each nearby label (neighboring) is weighted according to the strength of the link connecting it to the node in question. If there is a tie in the weighted neighbor count, the algorithm resolves it randomly.
\end{itemize}
 
 Given the stochastic nature of label propagation, different communities might end up the same label. To get rid of this issue, after the algorithm has been completed, a breadth-first search (BFS) is carried out on the subgraphs within each community. This step ensures that disconnected groups of nodes, those connected only through nodes of different communities in the original network, are correctly identified and separated, following the suggestion in \cite{raghavan2007near}.

\end{justify}
\section{Result and discussion}

We conducted a comprehensive investigation of the core-periphery structure and the community structure across the entire dataset of NIFTY 200 and NIFTY 500 indices. A dynamic analysis is performed treating the window of length 250 days and 1 day as sliding day. The finding of these analysis are discussed below as the following points;

\begin{itemize}

    \item Comparative study of the core-periphery structure on Random Matrix Theory.
    
    \item Comparative study of Community structure on Random Matrix Theory.

     \item  Portfolio optimization using community Information.
    
\end{itemize}

We elaborate on each of the above points below in detail.

\subsection{Comparative study of the core-periphery structure on random matrix theory}
\begin{justify}


To assess the efficiency of these correlation matrices, we compared the coreness values they produced against those generated by established algorithms, including Rossa, Rombach, and MINRES. To further evaluate their performance, we calculated the cosine similarity between the coreness values obtained from the full correlation method and those from the market and sector modes.

The cosine similarity \( S_c(\mathbf{X}, \mathbf{Y}) \) between two non-zero vectors \( \mathbf{X} \) and \( \mathbf{Y} \) is defined as:

\[
S_c(\mathbf{X}, \mathbf{Y}) := \frac{\mathbf{X^T} \mathbf{Y}}{\|\mathbf{X}\|_2 \|\mathbf{Y}\|_2}
\]

The cosine similarity ranges from -1 to 1, where -1 indicates complete opposition, 1 indicates identical vectors, and 0 indicates no correlation. Intermediate values represent varying degrees of similarity or dissimilarity between the vectors.


\begin{table}[ht]
\centering
\caption{Cosine similarity between the coreness values obtained through the full and mode correlation matrices from different core score estimation algorithms for the NIFTY 200 index.}

\begin{tabular}{|m{4cm}|m{4cm}|m{4cm}|}
\hline
\textbf{Methods} & \textbf{$C^{f}$ vs $C^{mar}$} & \textbf{$C^{f}$ vs $C^{sec}$} \\
\hline
Rossa & 0.59 & 0.44 \\
\hline
Rombach & 0.91 & 0.55 \\
\hline
MINRES & 0.63 & 0.63 \\
\hline
\end{tabular}
\label{tab:cosine_similarity_200}
\end{table}




\begin{table}[ht]
\centering
\caption{Cosine similarity between the coreness values obtained through the full and mode correlation matrices from different core score estimation algorithms for the NIFTY 500 index.}

\begin{tabular}{|m{4cm}|m{4cm}|m{4cm}|}
\hline
\textbf{Methods} & \textbf{$C^{f}$ vs $C^{mar}$} & \textbf{$C^{f}$ vs $C^{sec}$} \\
\hline
Rossa & 0.51 & 0.51 \\
\hline
Rombach & 0.84 & 0.67 \\
\hline
MINRES & 0.54 & 0.56 \\
\hline
\end{tabular}
\label{tab:cosine_similarity_500}
\end{table}


\begin{table}[ht]
\centering
\caption{Frobenius norm of the difference between the ideal core-periphery model and the normalized permuted adjacency matrices (\(\| \Delta_{\text{ideal}} - \Delta_0 \|_F\)) for different core score estimation algorithms on the NIFTY 200 index.}

\begin{tabular}{|m{3cm}|m{3cm}|m{3cm}|m{3cm}|}
\hline
\textbf{Mode} & \textbf{Rossa} & \textbf{Rombach} & \textbf{MINRES} \\
\hline
$C^{f}$ & 90.883 & 91.211 & 92.710 \\
\hline
$C^{mar}$ & 90.349 & 90.469 & 90.862 \\
\hline
$C^{sec}$ & 90.919 & 91.921 & 92.663 \\
\hline
\end{tabular}
\label{tab:difference_NIFTY 200}
\end{table}
\begin{table}[ht]
\centering
\caption{Frobenius norm of the difference between the ideal core-periphery model and the normalized permuted adjacency matrices (\(\| \Delta_{\text{ideal}} - \Delta_0 \|_F\)) for different core score estimation algorithms on the NIFTY 500 index.}

\begin{tabular}{|m{3cm}|m{3cm}|m{3cm}|m{3cm}|}
\hline
\textbf{Mode} & \textbf{Rossa} & \textbf{Rombach} & \textbf{MINRES} \\
\hline
$C^{f}$ & 204.469 & 204.835 & 206.145 \\
\hline
$C^{mar}$ & 204.116 & 204.142 & 204.517 \\
\hline
$C^{sec}$ & 204.400 & 204.624 & 206.427 \\
\hline
\end{tabular}
\label{tab:difference_NIFTY 500}
\end{table}

The tables \ref{tab:cosine_similarity_200} and \ref{tab:cosine_similarity_500} present the cosine similarity between coreness values obtained through full and mode correlation matrices across the core score estimation algorithms (Rossa, Rombach, and MINRES methods). For the NIFTY 200 Index, Table \ref{tab:cosine_similarity_200} shows that the Rombach method achieves the highest similarity, particularly between the full and market modes, with a value of 0.91. The Rossa method exhibits moderate performance, with a peak similarity score of 0.59 between the full and market modes. The MINRES method displays consistent performance across all modes, with a uniform similarity score of 0.63. In the NIFTY 500 Index, Table \ref{tab:cosine_similarity_500} reveals similar trends. The Rombach method continues to perform well, particularly between the full and market modes, with a similarity score of 0.84. The Rossa method again shows moderate performance, with its highest similarity score of 0.51 between the full and both market and sector modes. The MINRES method remains consistent, displaying moderate similarity across all modes with scores around 0.54.\\
To assess the efficacy of the full and mode correlation matrices, we followed the approach outlined below. First, we permuted the rows and columns of the ground truth adjacency matrices in the decreasing order of the core scores output by different methods and normalized the matrices so that their entries lie between 0 and 1. We denote the normalized permuted matrix as \( \Delta_0 \). We compare the ground truth adjacency matrix from the adjacency matrix of the ideal core-periphery structure\cite{Borgatti1999}. In the ideal core-periphery structure, core nodes are adjacent to other core nodes, core nodes are adjacent to periphery nodes, but periphery nodes are not adjacent to each other. The adjacency matrix corresponding to the idealized core-periphery structure is defined as:

\[
\Delta_{\text{ideal}} =
\begin{pmatrix}
\mathbf{1}_{k \times k} & \mathbf{1}_{k \times (N-k)} \\
\mathbf{1}_{(N-k) \times k} & \mathbf{0}_{(N-k) \times (N-k)}
\end{pmatrix},
\]

where \( \mathbf{1}_{m \times n} \) and \( \mathbf{0}_{m \times n} \) denote matrices filled with ones and zeros, respectively. In this model, \( k \) represents the number of core nodes, and we set \( k \) to \( N/4 \) for both data sets.

For comparison, we computed the frobenius norm difference between the normalized permuted matrix \( \Delta_0 \) and the idealized core-periphery matrix \( \Delta_{\text{ideal}} \), which is defined as:

\[
\|\Delta_0 - \Delta_{\text{ideal}}\|_F = \sqrt{\sum_{i=1}^{m} \sum_{j=1}^{n} |\Delta_{0_{ij}} - \Delta_{\text{ideal}_{ij}}|^2}
\]

where \( \Delta_{0_{ij}} \) and \( \Delta_{\text{ideal}_{ij}} \) are the \((i,j)\)th elements of \( \Delta_0 \) and \( \Delta_{\text{ideal}} \), respectively.

Tables \ref{tab:difference_NIFTY 200} and \ref{tab:difference_NIFTY 500} provide the frobenius norm of the difference between the ideal core-periphery structure and the normalized permuted adjacency matrices for different core score estimation algorithms (Rossa, Rombach, MINRES) in the NIFTY 200 and NIFTY 500 indices. For the NIFTY 200 index, the market mode (\(C^{mar}\)) consistently shows the smallest frobenius norm across all algorithms, indicating the closest alignment with the ideal core-periphery structure. The full mode (\(C^{f}\)) and sector mode (\(C^{sec}\)) exhibit larger deviations. Similarly, for the NIFTY 500 index, the market mode (\(C^{mar}\)) again demonstrates the smallest frobenius norm. The full (\(C^{f}\)) and sector (\(C^{sec}\)) modes show larger deviations from the ideal structure. Overall, the market mode consistently outperforms the full and sector modes in terms of proximity to the ideal core-periphery structure, making it the most effective mode for core-periphery analysis in both indices.


We also analyzed the dynamical stability of the obtained core-periphery structure over time using the full and RMT-based correlation matrices of the NIFTY 200 and NIFTY 500 indices. We constructed PMFG-filtered networks using a one-day rolling window from the full, market, and sector modes. Next, we computed the core-periphery centralization (cp-centralization) \(C^{cp}\) for each rolling window, following the method proposed by Rossa et al.\cite{Rossa2013}. The core-periphery centralization \(C^{cp}\) is defined as:

\[
C^{cp} = 1 - \frac{2}{n-2} \sum_{k=1}^{n-1} \alpha_k,
\]

where \((\alpha_1, \alpha_2, \ldots, \alpha_{n-1})\) denotes the core-periphery profile of the network. A high value of \(C^{cp}\) suggests a pronounced core-periphery structure in the network, whereas a lower value indicates a more randomized structure with no distinct core-periphery structure. Figures \ref{fig:cp_centralization_nifty200_500}(a) and \ref{fig:cp_centralization_nifty200_500}(b) display the heat map of cp-centralization across all windows from the NIFTY 200 and NIFTY 500 indices. This plot shows that Market mode consistently detects a strong core-periphery structure over the full and sector modes.

To assess the significance of the observed core-periphery structure, we computed cp-centralization value \( C^{cp} \), we use a simulation approach involving the generation of 100 randomized networks for each original network. These randomized networks are constructed to maintain the same degree distribution as the original ones\cite{hakimi1962realizability, kleitman1973algorithms}. For each randomized network, we calculate the cp centralization value, denoted \( C^{cp}_{i,rand} \) for \( i = 1, 2, \ldots, 100 \). The \( p \)-value is then computed as the fraction of randomized networks with a cp-centralization value greater than the observed \( C^{cp} \), given by \( p = \frac{\#\{i : C^{cp}_{i,rand} > C^{cp}\}}{100} \). The following hypotheses are tested:

\textbf{Null Hypothesis} (\( H_0 \)): The observed cp-centralization \( C^{cp} \) is not statistically significant and could be due to random chance.

\textbf{Alternative Hypothesis} (\( H_a \)): The observed cp-centralization \( C^{cp} \) is statistically significant, indicating that it is unlikely to have occurred by chance.

A small \( p \)-value (usually less than 0.01 or 0.05) leads to rejecting the null hypothesis \( H_0 \) and accepting the alternative hypothesis \( H_a \), thereby concluding that the observed cp-centralization \( C^{cp} \) reflects a significant core-periphery structure rather than random fluctuation.


\begin{figure}[htbp]
    \centering
    \begin{subfigure}{0.9\textwidth}
        \centering
        \includegraphics[width=\textwidth]{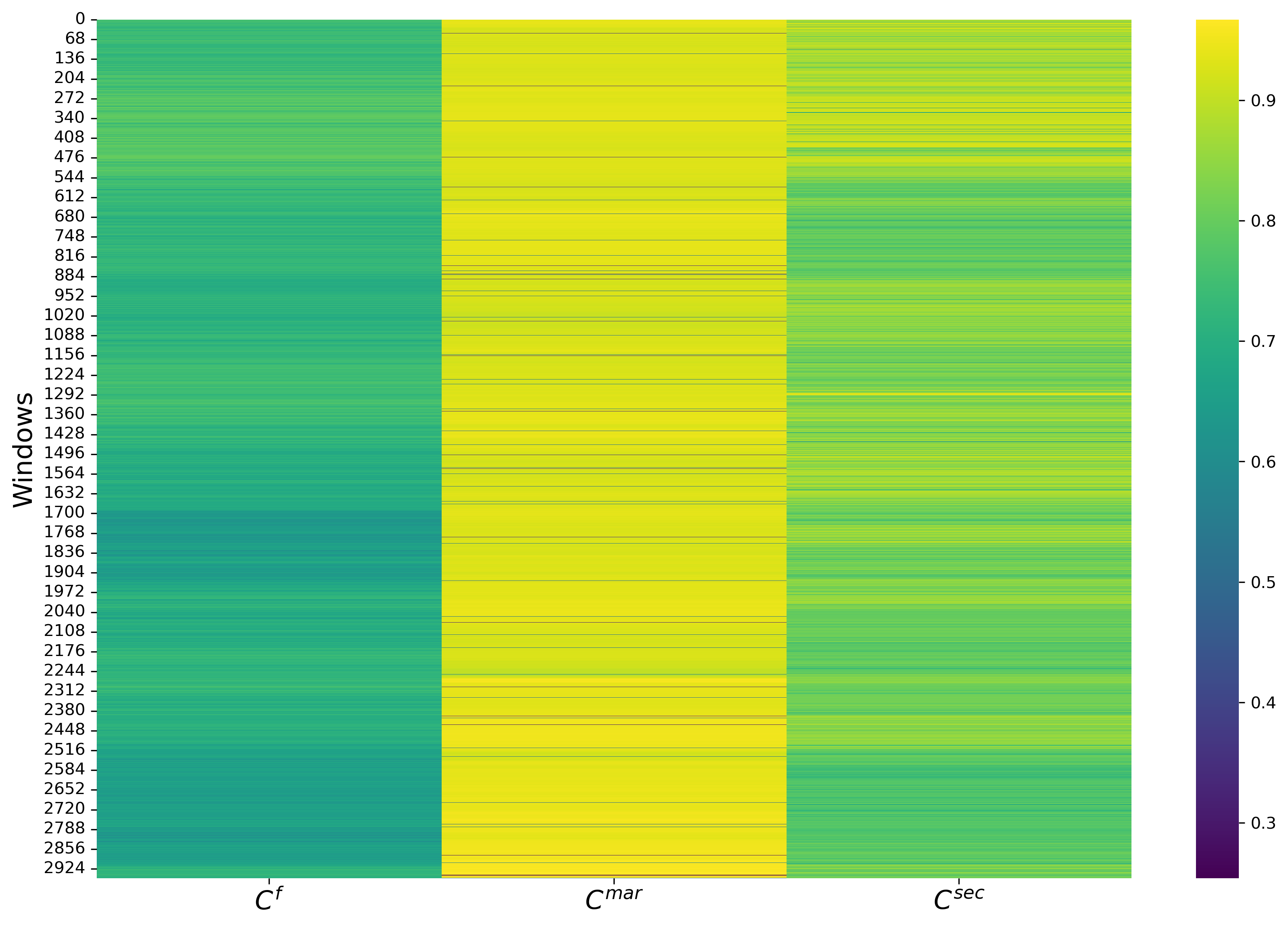}
        \textbf{(a)} 
        \label{fig:nifty200_heatmap_output}
    \end{subfigure}
    
    \vspace{0.5cm} 
    
    \begin{subfigure}{0.9\textwidth}
        \centering
        \includegraphics[width=\textwidth]{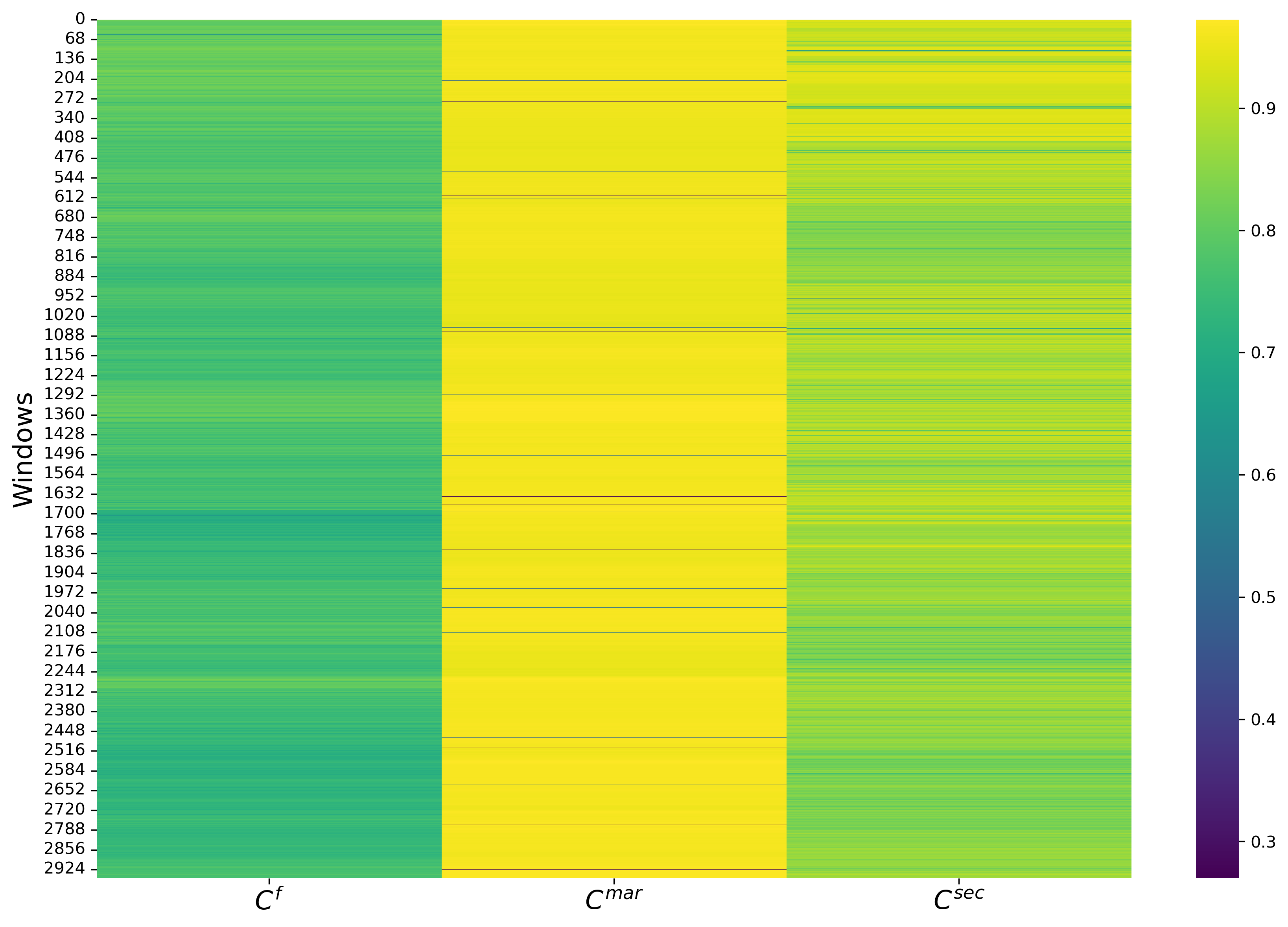}
        \textbf{(b)} 
        \label{fig:nifty500_heatmap_output}
    \end{subfigure}
    
    \caption{The cp-centralization $\mathbf{C^{cp}}$ over time for the (a) NIFTY 200 and (b) NIFTY 500 indices.}
    \label{fig:cp_centralization_nifty200_500}
\end{figure}


In our analysis, we evaluated statistical testing across three methods: Full cross-correlation, market mode, and sector mode, using one-day rolling windows over two datasets: the NIFTY 200 index and the NIFTY 500 index. For the NIFTY 200 index, spanning 2958 rolling windows from january 2010 to december 2022, the results were as follows: Full cross-correlation yielded 99.73\% of windows at the 1\% significance level and 99.80\% at the 5\% level; market mode showed 97.94\% and 99.80\%; and sector mode showed 98.95\% and 99.80\%, respectively. For the NIFTY 500 index, with 2954 rolling windows, the results were as follows: Full cross-correlation had 99.97\% at the 1\% significance level and 99.97\% at the 5\% level; sector mode had 99.56\% at both significance levels; and market mode had 98.58\% at both significance levels.

\end{justify}
\subsection{Comparative study of community structure on random matrix theory}
\begin{justify}

In this section, we performed a comparative community analysis on the PMFG networks ($C^{f}$), ($C^{sec}$) and ($C^{mar}$), following a windows-wise dynamic approach and also over the entire datasets. We used community detection algorithms such as the Louvain Algorithm, the Asynchronous Label Propagation Algorithm, and the Newman Girvan to capture the community structure. First, we provide the detected modularity value for each network communities in the Table \ref{tab:Modularity}. The figure indicates that the modularity of the sector mode is close to full cross correlation and even greater in some of the cases for both the datasets, indicating that it capture significant community structure. Thus the community patterns of the market can be studied in the sector mode not in the market mode. Note that in the market mode, modularity is close to zero and even negative, this might be due to less correlated set of nodes, leading to weaker community structure. A similar observation can be seen in Table \ref{tab:Modularity1} for the case of NIFTY 200.  
\begin{table}[ht]
\centering
\caption{Modularity values obtained over the PMFG networks corresponding to the full cross-correlation, market mode and the sector mode via different community detection algorithms for the NIFTY 500 index.}

\begin{tabular}{|m{3cm}|m{3cm}|m{3cm}|m{3cm}|}
\hline
\textbf{Methods} & \textbf{$C^{f}$} & \textbf{$C^{mar}$ } & \textbf{$C^{sec}$} \\
\hline
Louvain &  0.716 & 0.178 & 0.721 \\
\hline
Label propagation  & 0.641 & 6.662e-16 & 0.652 \\
\hline
Newman Girvan  & 0.510 & -1.48e-06 & 0.340 \\
\hline
\end{tabular}
\label{tab:Modularity}
\end{table}

\begin{table}[ht]
\centering
\caption{Modularity values obtained over the PMFG graph corresponding to the full cross-correlation, market mode and the sector mode via different community detect algorithms for the NIFTY 200 index.}

\begin{tabular}{|m{3cm}|m{3cm}|m{3cm}|m{3cm}|}
\hline
\textbf{Methods} & \textbf{$C^{f}$} & \textbf{$C^{mar}$ } & \textbf{ $C^{sec}$} \\
\hline
Louvain  &  0.662 & 0.160 & 0.658 \\
\hline
Label propagation  & 0.630 & -2.220e-16 & 0.660 \\
\hline
Newman Girvan & 0.496 & -7.304e-06 & 0.497 \\
\hline
\end{tabular}
\label{tab:Modularity1}
\end{table}

A similar pattern could be seen over window-wise dynamic analysis treating the 250 days as window size and one day as sliding. So, there are a total of 2958 windows in NIFTY 500 and 2954 in NIFTY 200 datasets. In figures \ref{fig:combined_modularity_200} (a), \ref{fig:combined_modularity_200} (b), and \ref{fig:combined_modularity_200} (c), it is clearly visible that the modularity of the sector mode is consistently higher than the full cross-correlation across all windows for the datasets. We have also seen the dynamics of the number of communities as provided in Fig. \ref{fig:modularity3}. It is seen that the number of communities inside the market mode is larger than the full cross-correlation and is further larger than the sector mode.  The market mode captures broad market-wide trends that tend to affect all stocks to some extent. Because of this, the correlations between stocks in this mode are generally weaker and more homogeneous compared to the full cross-correlation matrix, while sector mode emphasizes correlations within specific sectors,  and then strong intra-sector correlations in sector mode result in a higher modularity score for fewer, larger communities.


\begin{figure}[htbp]
    \centering
    \begin{subfigure}{0.7\textwidth}
        \centering
        \includegraphics[width=\textwidth]{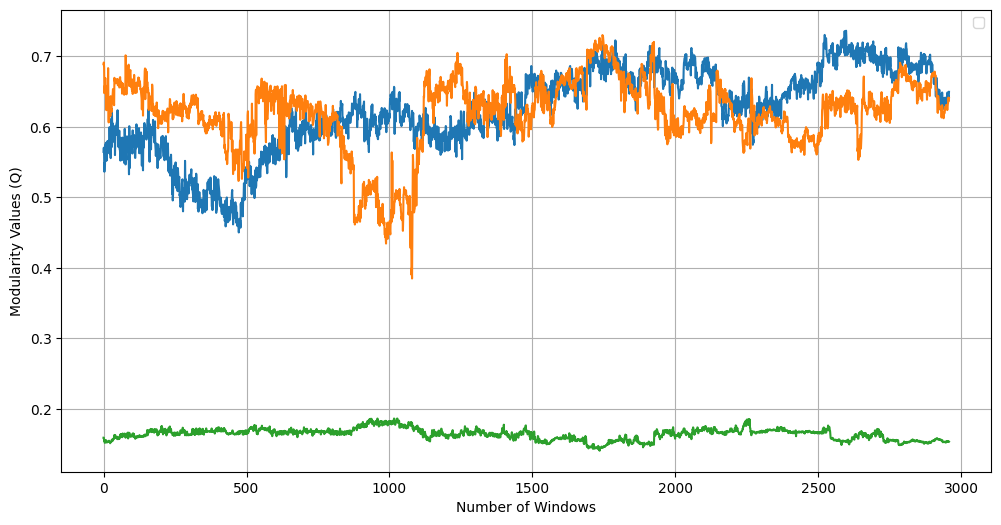}
        \caption{Louvain}
        \label{fig:modularity}
    \end{subfigure}
    
    \vspace{0.5cm} 
    
    \begin{subfigure}{0.7\textwidth}
        \centering
        \includegraphics[width=\textwidth]{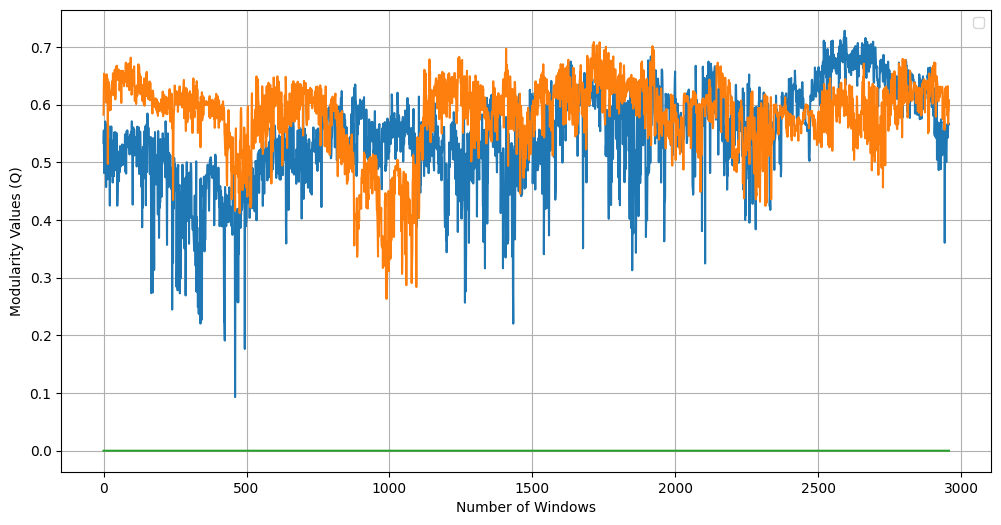}
        \caption{Label Propagation}
        \label{fig:modularity1}
    \end{subfigure}

    \vspace{0.5cm} 
    
    \begin{subfigure}{0.7\textwidth}
        \centering
        \includegraphics[width=\textwidth]{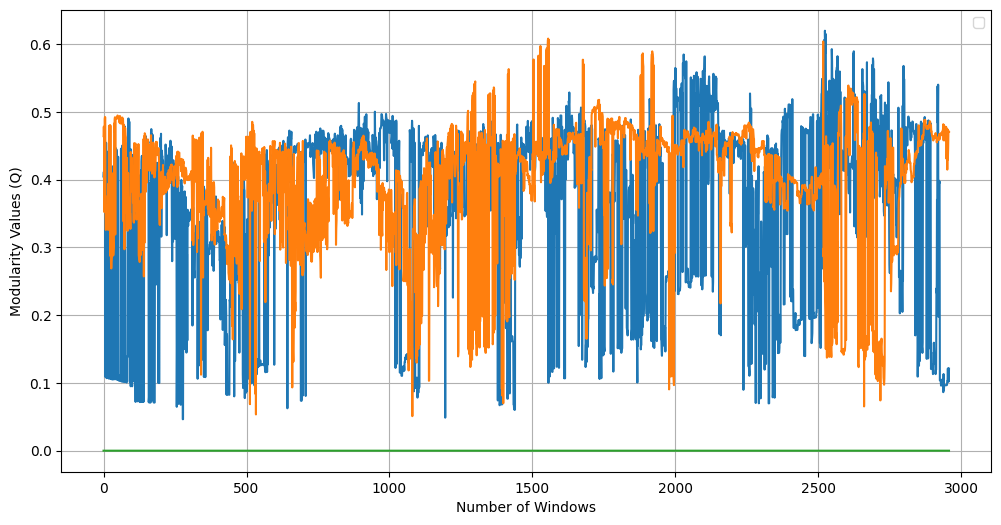}
        \caption{Newman}
        \label{fig:modularity2}
    \end{subfigure}

    \caption{The dynamics of modularity series generated on networks based on the full cross-correlation matrix (blue), sector mode (orange), and market mode (green) over the NIFTY 200 Index, using three different methods: (a) Louvain, (b) Label Propagation, and (c) Newman.}
    \label{fig:combined_modularity_200}
\end{figure}


\begin{figure}[h!]
    \centering
    \includegraphics[width=0.9\textwidth]{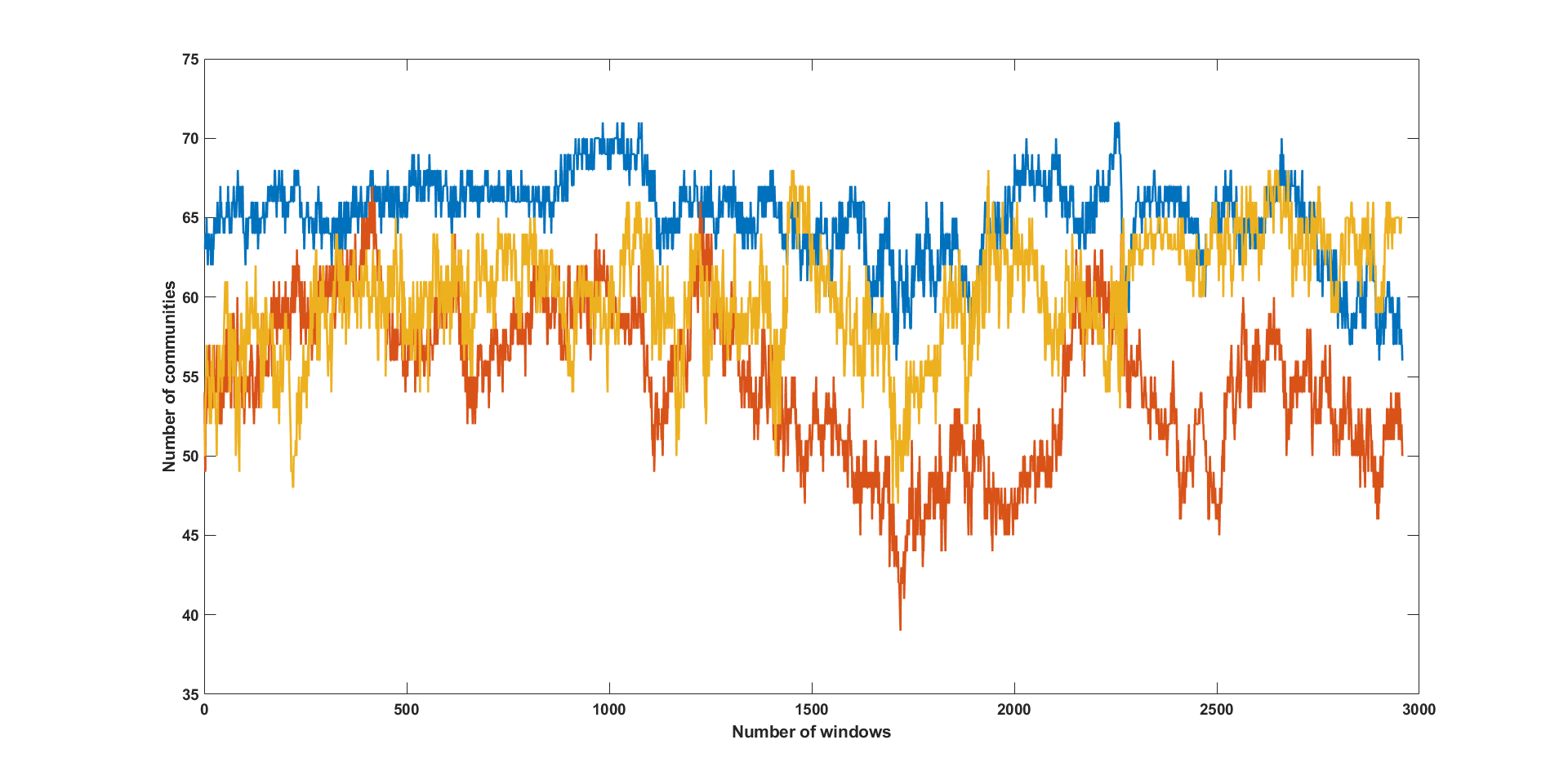}
\caption{The dynamics of the number of Louvain communities over the windows of the NIFTY 200 Index. The blue curve represents the number of communities of the market mode, the yellow curve is for the full cross-correlation matrix, while red gives the number of communities of sector mode per window.}

    \label{fig:modularity3}
\end{figure}

\end{justify}

\begin{figure}[h!]
    \centering
    \includegraphics[width=\textwidth]{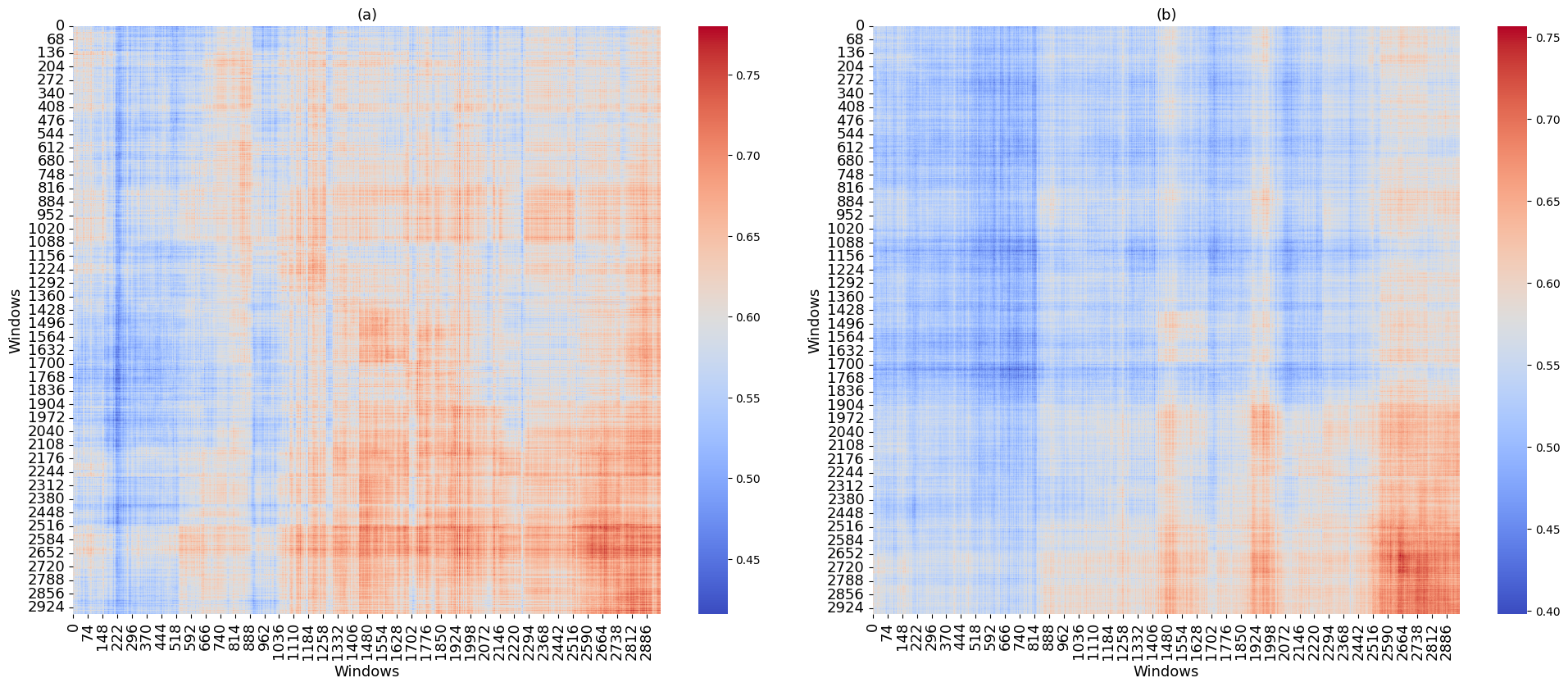}
\caption{The dynamics of the NMI between the Louvain communities corresponding to the sector mode and full cross-correlation a) NIFTY 200 and b) NIFTY 500.}

    \label{fig:nmi_dynamics}
\end{figure}

\begin{justify}
On the other hand, we have also checked the similarity between the communities of the network $C^{f}$, with the communities of the network $C^{sec}$ for the three community detection algorithms using the normalized mutual information metric.  This is also performed over the entire dataset and window-wise following the same process as done for the modularity detection case. For the entire dataset, the results are provided in Tables \ref{tab:NMI500}. The second column in this table was generated on the NIFTY200 dataset and the third column on the NIFTY500 dataset. For the case of Louvain and Label propagation algorithms, the NMI scores are quite good.  This implies that the sector mode preserves the communities of the network based on a full cross-correlation matrix. In addition, we created a heat map based on the dynamics of the NMI computed between the Louvain communities of the network based on sector mode and full cross-correlation, in each window of both data sets provided in Fig. \ref{fig:nmi_dynamics}. Note that the NMI values are greater than 0.45 and less than 0.80. Results over NIFTY500 are provided in the supplementary information.

\end{justify}

\begin{table}[ht]
\centering
\caption{NMI values for the communities of the sector and market mode compared with the communities obtained over the PMFG graph corresponding to the full correlation for the NIFTY 500 index.}
\begin{tabular}{|m{3cm}|m{3cm}|m{3cm}|}
\hline
\textbf{Methods} & \textbf{$C^{f}$ vs $C^{sec}$} & \textbf{$C^{f}$ vs $C^{mar}$} \\ 
\hline
Louvain & 0.660 & 0.706 \\
\hline
Label Propagation & 0.682 & 0.703 \\
\hline
Newman Girvan & 0.526 & 0.562 \\
\hline
\end{tabular}
\label{tab:NMI500}
\end{table}

\subsection{Portfolio optimization using community information}

\begin{justify}
 
This portfolio construction analysis is performed for the entire 13-year period for the data sets of the NIFTY 200 and NIFTY 500 indices from the National Stock Exchange of India. Notably, during this period, the market experienced a significant downturn in 2020 due to the COVID-19 pandemic, resulting in prolonged volatility. Our analysis utilizes moving windows of 250 days (equivalent to 1 year). For each day $t$ within this window, we construct the PMFG-filtered network based on the full cross-correlation matrix ($C^{f}$), the market mode ($C^{mar}$), and the sector mode ($C^{sec}$) derived from the previous 250 days. We then assess the stability of the portfolio over the following 250 days, encompassing holding periods ranging from 1 to 250 market days. Since Louvain community detection performs well in our data set to capture the significant community structure, we further used this algorithm to apply it to PMFG networks. we used these communities for portfolio construction following the different strategies for stock selection. Then compare different investment strategies by assigning weights using the equal weighting method and the Markowitz method, as described below. The strategies for choosing the stocks from the communities for the portfolio construction are as follows; 

\

$\mathbf{P^f_{{max}}}$: This notion represents the portfolio constructed over the PMFG network corresponding to the full cross-correlation matrix. One stock with having maximum Sharpe ratio was selected from each of the communities. 

\

$\mathbf{P^f_{{rand}}}$: This portfolio is constructed by choosing one stock randomly from each community of the network constructed over the full cross-correlation network.





\

$\mathbf{P^{sec}_{{max}}}$: This portfolio was constructed using the same strategy as the portfolio $\mathbf{P^f_{{max}}}$ but in sector mode.

\

$\mathbf{P^{sec}_{{ran}}}$:  This portfolio constructed using the same strategy of the portfolio $\mathbf{P^f_{{rand}}}$ but over the sector mode.

\

$\mathbf{P^{{MKT}}}$: This is the "market portfolio"
that comprises all stocks in the respective market (140 in NIFTY 200 and 314 in NIFTY 500).

\

Here, we wish to point out that the size of the portfolio constructed in each four strategies can be seen in Fig. \ref{fig:modularity3} across all windows of the NIFTY200 dataset.
The network based on full cross-correlation matrix includes all the relationships between stocks, capturing both sectoral (intra-sector) and cross-sector (inter-sector) correlations. Through this, we will get a more holistic understanding of the overall dynamics of the market. Thus, considering the full cross-correlation matrix, the portfolio is not limited to within-sector correlations. This enables better diversification across different sectors and market segments, reducing overall portfolio risk. This also captures broad market trends that may not be evident when focusing only on sector-specific mode or market mode. This helps in selecting stocks that are not only strong within their sectors, but also well-positioned within the broader market context.\\

\begin{figure}[h!]
    \centering
    \includegraphics[width=\textwidth]{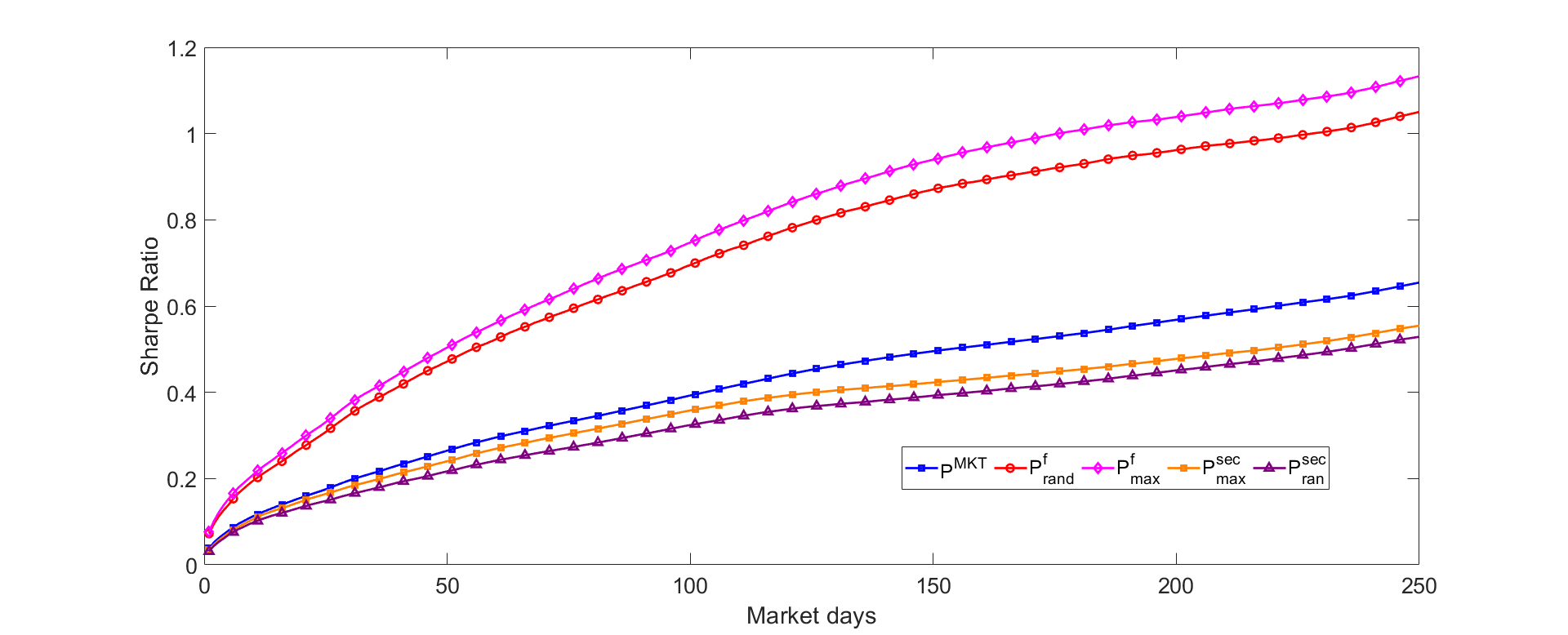}
\caption{ The plot reports the Sharpe ratio comparison between the $\mathbf{P^f_{{max}}}$, $\mathbf{P^f_{{rand}}}$, $\mathbf{P^{mar}_{{max}}}$, $\mathbf{P^{mar}_{{ran}}}$, $\mathbf{P^{sec}_{{max}}}$, $\mathbf{P^{sec}_{{ran}}}$, and $\mathbf{P^{{MKT}}}$, respectively, for holding periods ranging from 1 to 250 days for the NIFTY 200 index; weights assigned with uniform method.
}

    \label{fig:shrp_ratio_results_uniform_200}
\end{figure}

\begin{figure}[h!]
    \centering
    \includegraphics[width=\textwidth]{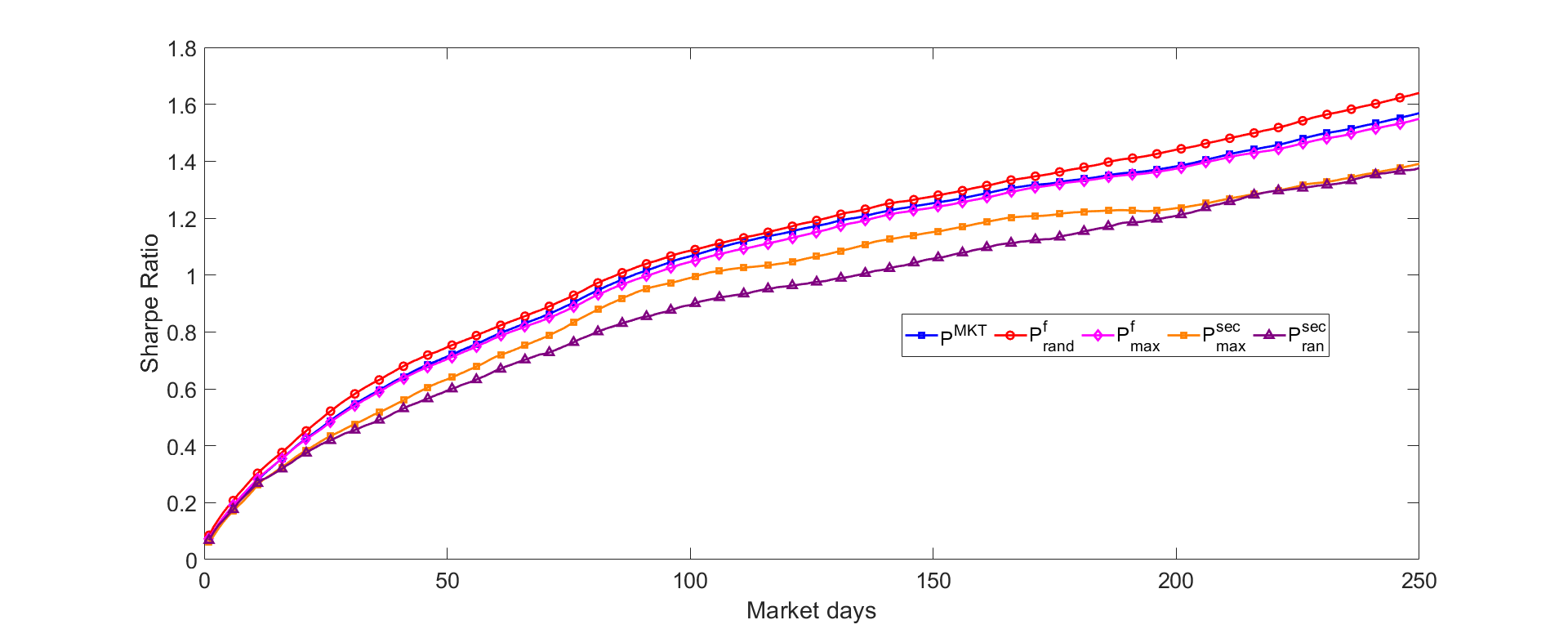}
\caption{The plot reports the Sharpe ratio comparison between the $\mathbf{P^f_{{max}}}$, $\mathbf{P^f_{{rand}}}$, $\mathbf{P^{mar}_{{max}}}$, $\mathbf{P^{mar}_{{ran}}}$, $\mathbf{P^{sec}_{{max}}}$, $\mathbf{P^{sec}_{{ran}}}$, and $\mathbf{P^{{MKT}}}$, respectively, for holding periods ranging from 1 to 250 days for the NIFTY 200 index; weights assigned with Markowitz method.}

    \label{fig:shrp_ratio_results_markowitz_200}
\end{figure}


This market mode represents overall market behaviour, which often dominates individual sector movements. When constructing a portfolio based on the market mode, the portfolio effectively aligns with the general market trend, which could be more stable and less volatile than individual sectors. On the other hand, the sector mode captures intra-sector dynamics, and it might miss out on the broader inter-sector relationships and market-wide trends. sectors can sometimes be influenced by specific localized factors, which may not reflect the general direction of the market, leading to a poor performance relative to a more diversified approach.

From supplementary figures we observe that both the uniform and Markowitz method schemes yield similar results for the NIFTY500.

\end{justify}

\section{Conclusion}

\begin{justify}

In this study, we have comprehensively explored the core-periphery and the community structures within financial networks using Random Matrix Theory (RMT). First, we applied the RMT to filter out the significant correlation structure present in the financial market data. Then we constructed the network over the obtained structure in order to study the important aspects such as the core-periphery and community structure. Our findings suggest that the core-periphery structure is primarily captured by the market mode, while the sector mode effectively captures community structures. We also find that the correlation between the coreness vectors of the core-score estimation algorithms based on the full- and market-mode correlation matrices is high. Thus, it outperforms both the full correlation matrix and the sector mode in identifying the ideal core-periphery structure, as indicated by lower frobenius norms. Further, the modularity score of the communities corresponding to the sector mode is higher than the communities corresponding to the full cross-correlation and the market mode networks. Hence, the sector mode has a more significant community structure than the full correlation matrix and market mode. In support of this, the NMI between the communities of the full cross-correlation network and the sector-mode-based network is also high, indicating that the sector-mode community structure is closer to the full cross-correlation network structure. Furthermore, the dynamic analysis also supports that the market mode consistently maintains a strong core-periphery structure over the 13 years, while the sector mode excels in capturing community structure. 

Next, we utilized the communities obtained from the networks based on full cross-correlation and sector mode to construct the risk-diversified portfolios. Our analysis showed that portfolios containing inter-community stocks constructed using the full correlation matrix outperform portfolios based on the sector mode. Additionally, our analysis shows that portfolios containing inter-community stocks outperform market portfolios (i.e.; investing in all stocks) in both the uniform and Markowitz methods. In conclusion, this research demonstrates the utility of full and RMT-based correlation matrices in enhancing our understanding of financial markets, particularly in core-periphery and community structures. In the future, RMT-based correlation matrices could be valuable to understand various systems beyond financial networks, such as social, biological, and transportation networks.

\end{justify}

\section*{Acknowledgments}
\begin{justify}
We thank the Shiv Nadar Institution of Eminence for providing the computational facilities and
the necessary infrastructure needed to carry out the present research. 
\end{justify}

%
%
%

\bibliographystyle{elsarticle-num} 
\bibliography{References}






\end{document}